\documentclass[12pt]{article}   

\usepackage{opex3} 
\usepackage[draft]{hyperref} 
\usepackage{amsmath}
\usepackage{epsfig}
\newcommand{\nm}{~{\text{nm}}}
\newcommand{\f}{ f }
\newcommand{\mum}{~\mu{\text{m}}}

\newcommand{\Rneff}{n_i^{-1}}
\newcommand{\etaPL}{\eta_{\text{PL}}}
\newcommand{\etaPLm}[1]{\eta_{\text{PL},#1}}

\newcommand{\Real}[1]{\text{Re}\left\{ #1 \right\}}

\newcommand{\Bmm}{B_mB^*_{m'}}

\begin{document}


\title{Hybrid gap modes induced by fiber taper waveguides: application in spectroscopy of
single solid-state emitters deposited on thin films.}


\author{Marcelo Davan\c co$^{* 1,2}$ and Kartik Srinivasan$^1$}
\address{$^1$Center for Nanoscale Science and Technology,
National Institute of Standards and Technology, Gaithersburg, MD\\
$^2$Maryland NanoCenter, University of Maryland, College Park, MD,
20742\\$^*$Corresponding author: mdavanco@nist.gov}

\begin{abstract*}
We show, via simulations, that an optical fiber taper waveguide can
be an efficient tool for photoluminescence and resonant, extinction
spectroscopy of single emitters, such as molecules or colloidal
quantum dots, deposited on the surface of a thin dielectric
membrane. Placed over a high refractive index membrane, a tapered
fiber waveguide induces the formation of hybrid mode waves, akin to
dielectric slotted waveguide modes, that provide strong field
confinement in the low index gap region. The availability of such
gap-confined waves yields potentially high spontaneous emission
enhancement factors ($\approx20$), fluorescence collection
efficiencies ($\approx23~\%$), and transmission extinction
($\approx20~\%$) levels.  A factor of two improvement in
fluorescence and extinction levels is predicted if the membrane is
instead replaced with a suspended channel waveguide. Two
configurations, for operation in the visible ($\approx 600 \nm$) and
near-infrared ($\approx 1300 \nm$) spectral ranges are evaluated,
presenting similar performances.
\end{abstract*}

\vskip 0.5 cm







\section{Introduction}

In Ref.~\cite{Almeida.OL.04}, a dielectric waveguide structure,
consisting of two high refractive index regions separated by a
narrow, low refractive index slot, was shown to support propagating
modes with a very high field concentration in the slot region. The
strong field confinement was shown to be related to the continuity
of the electric displacement vector component across the gap, and to
increase with index contrast. The availability of such modes offers
numerous possibilities for the realization of integrated optical
devices that exploit spatial localization~\cite{ref:Yang_AHJ},
enhanced nonlinearities~\cite{ref:Koos}, and field-matter
interaction~\cite{ref:Jun_YC} within the gap region.

In this paper, we use electromagnetic simulations to show that an
optical fiber taper waveguide (sometimes called a micro- or
nanofiber waveguide) can be used as an efficient probe for resonant
and non-resonant spectroscopy of individual emitters bound to the
surface of thin dielectric membranes. High probing efficiency is
possible due to the availability of hybrid guided waves akin to the
air-slot waveguide modes introduced in~\cite{Almeida.OL.04}. Such
hybrid waves, referred to as gap modes, display strong field
concentration in the gap between fiber and membrane and at the same
time couple efficiently to the access optical fiber waveguide mode.
The first feature leads to spontaneous emission rate enhancement,
while the second, combined with high percentage coupling of
spontaneous emission into gap modes, leads to high fluorescence
collection efficiency.

The ability to perform spectroscopy of individual guest emitters in
solid state hosts is desirable not only from a materials science
perspective, in which the microscopic properties of either the host
or the composite are investigated~\cite{ref:Moerner_review}, but
also for experiments in quantum optics, in which the control and
manipulation of the quantum states of individual guest emitters is
sought~\cite{ref:Moerner,ref:Sandoghdar_nature_460}. Essential for
both single emitter spectroscopy and quantum state manipulation is
the availability of highly efficient access channels to the dipole,
which, our previous work
suggests~\cite{srinivasan:091102,ref:Davanco,Davanco2}, can be
provided by an optical fiber taper waveguide.

\begin{figure}[t]
\centerline{\includegraphics[width=12.0cm]{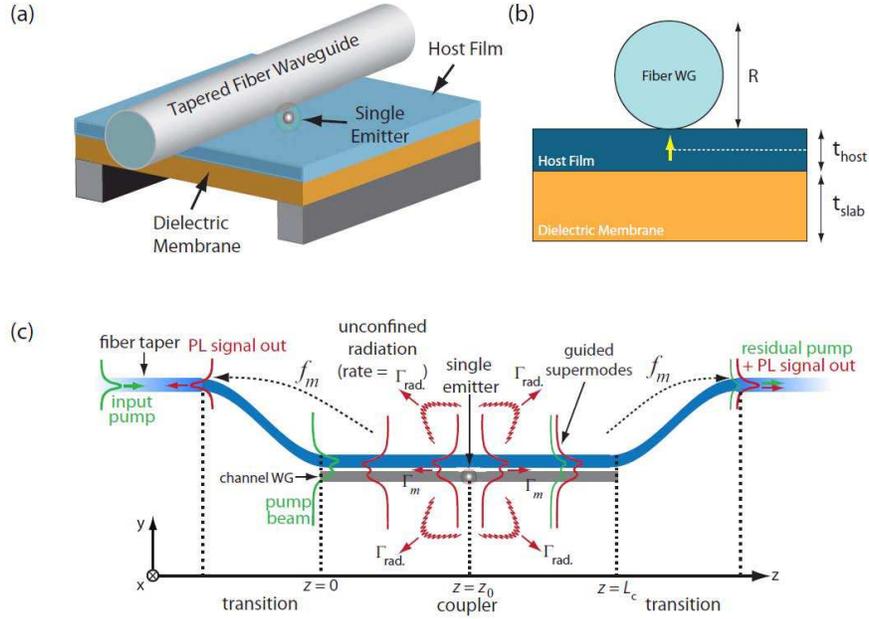}}
\caption{(a) Tapered fiber waveguide-based probing configuration for
emitters deposited on the surface of a dielectric membrane. An
individual emitter, embedded in a host thin film, is depicted under
the fiber. (b) Cross-section of structure in (a). (c) Schematic of
single emitter excitation and PL collection via the tapered fiber
probe. A non-resonant pump signal is injected into the input fiber
and converted into a guided supermode of the composite waveguide,
illuminating the slab-embedded dipole. The dipole radiates into
guided and radiative supermodes, with rates $\Gamma_m$ and
$\Gamma_{rad}$ respectively. Power is transferred with efficiency
$\f_m$ from the
  supermode to the fiber mode and vice-versa. } \label{fig-probing-scheme}
\end{figure}

The optical fiber taper waveguide is a single mode optical fiber
whose diameter is adiabatically and symmetrically reduced to a
wavelength-scale minimum, resulting in a low-loss, double-ended
device with standard fiber input and output. The manner in which
emitters on the surface of a dielectric membrane may be optically
accessed through the fiber is shown in
Figs.~\ref{fig-probing-scheme}(a)-(c). The wavelength-scale single
mode region of the optical fiber waveguide is brought into proximity
with the top surface of the membrane, over a length of several
wavelengths. Fiber and membrane together form the composite
dielectric waveguide with cross-section shown in
Fig.~\ref{fig-probing-scheme}(b), which supports a complete set of
guided, leaky, and radiation supermodes originating from the
hybridization of fiber and slab modes~\cite{ref:Snyder_Love} (note
that gap modes are indeed supermodes of the hybrid fiber-slab
waveguide, as discussed in Section~\ref{section 3}). As illustrated
in Fig.~\ref{fig-probing-scheme}(c), for non-resonant
photoluminescence (PL) spectroscopy measurements, part of the
non-resonant pump power initially carried by the fiber is coupled to
supermode waves that reach the emitter. Illuminated by the pump, the
emitter radiates (at a red-shifted wavelength) into supermodes of
the composite waveguide, and a portion of the radiated power is
outcoupled through the two fiber ends. Using the same method, we
also envision the possibility of resonantly exciting the dipole, in
which case the radiated fields (resonance fluorescence) would be
coherent and at the same wavelength as the pump signal.
Backward-propagating resonance fluorescence can potentially be
collected from the input fiber port by way of a directional coupler.
In the forward direction, the power at the output fiber is given by
the interference between the excitation and resonant fluorescence
signals, and may result in an enhanced or diminished transmission
level~\cite{Davanco2,gerhardt.prl.033601}.

In~\cite{ref:Davanco}, we analyzed the collection efficiency of the
fiber taper waveguide for light emitted by a dipole embedded in a
high refractive index membrane. We showed that the collection
efficiency, achieved by tapping into emitted slab-confined waves, is
potentially highly superior to that attainable with standard
free-space collection using high numerical aperture optics. While
the existence of the aforementioned gap modes was mentioned in that
article, their potential application in performing efficient
spectroscopy of individual surface-bound dipoles was only briefly
discussed, and not analyzed at any level of detail. Such an analysis
is accomplished in the following sections.

This paper is organized as follows. In Section~\ref{section 1}, our
simulation model and methods are explained. In Section~\ref{section
2}, simulation results of fluorescence collection for two
configurations appropriate for visible and infrared wavelengths are
given. Section~\ref{section 3} analyzes the collection efficiency
results in terms of a hybrid waveguide mode decomposition.
Parameters from this analysis are used in Section~\ref{section 4} to
show the possibility of resonant field extinction by a single
dipole. A discussion of the results follows in Section~\ref{section
5}, and Section~\ref{section 6} concludes the paper.

\section{Model and methods}
\label{section 1} We envision a general spectroscopy configuration
for probing individual emitters in highly dilute guest-host material
systems, examples of which are: colloidal quantum dots (CdSe,
CdSe/ZnO, PbS, PbSe, etc) in polymer~\cite{ref:Lee_Bawendi} or
sol-gel~\cite{ref:Schaller} hosts; organic dyes in
polymer~\cite{ref:zumbush} or small-molecule crystalline
hosts~\cite{ref:Moerner,ref:orrit,ref:Harms}; rare-earth ions in
transparent solid-state hosts\cite{ref:bottger}. In our
configuration, a thin film of the guest-host material system would
be produced on top of a dielectric, high refractive index material,
chosen appropriately according to the guest emitter transition
wavelength ranges. For instance, a Si ($n \approx3.5$) membrane
could be used for near-infrared wavelengths above $1\mum$, while
SiN$_x$ ($n \approx 2.0$) could be used for visible light emission.
We point out that such thin-film structures may be produced in most
cases with well-known, standard nanofabrication techniques. As
depicted in Figs.\ref{fig-probing-scheme}(a) and (b), a tapered
optical fiber waveguide brought into contact with the host material
provides both the excitation and collection channels to the guest
emitters. We analyze this structure with the same method as
in~\cite{ref:Davanco}, where single emitter collection efficiency
from dipoles embedded in a dielectric membrane was studied. The
simulation model and analysis methods are briefly described below.

\subsection{Simulation model}
We model the problem as in Fig.\ref{fig-probing-scheme}(b). An
individual electric point dipole embedded in a thin host film on the
surface of a dielectric membrane of thickness $t_\text{slab}$ and
refractive index $n_\text{slab}$ is probed by an optical fiber taper
waveguide of radius $R$ and index $n_\text{fiber}=1.45$. The dipole,
oriented normal to the membrane surface, is assumed to be at the
center of a dielectric host film of index $n_\text{host}$ and
thickness $t_\text{host}$. It is also assumed to be aligned with the
center of the probing fiber. As in Fig.~\ref{fig-probing-scheme}(c),
an excitation signal, resonant or non-resonant with one of the
emitter's transitions, is launched into the fiber input and
adiabatically reduced in size as the fiber is tapered, exciting
supermodes of the coupler structure. Supermodes with sufficient
lateral confinement illuminate the dipole, at a position $z=z_0$
along the coupler. Under non-resonant excitation, the dipole emits
coupler supermodes in the $\pm z$ directions, at a red-shifted
wavelength. The emitted supermodes are converted into input and
output fiber modes through the taper transition regions, after which
emission is detected.

\subsection{Fluorescence collection simulation} \label{subsection:simulation}To estimate the PL collection
efficiency of our fiber-based probing scheme, we simulated a single
classical electric dipole radiating in the composite dielectric
waveguide of Fig.~\ref{fig-probing-scheme}(a), using the Finite
Difference Time Domain (FDTD) method~\cite{ref:Davanco}. The
simulation provided the steady-state fields over the entire
computational window, which
was cubic, with more than six wavelengths in size. These were used to calculate an upper bound
for the percentage of the total emitted power $P_{\text{Tot.}}$
coupled to the fundamental optical fiber mode at an arbitrary
position $z$ along the guide, with the expression
\begin{equation}
\eta_{\text{PL}} = 2\frac{P_z}{P_{\text{Tot.}}}f_{\text{fiber}}.
\label{eq_eta_PL_fdtd}
\end{equation}
Here, $P_z$ is the power flowing normally through the constant-$z$
plane, $f_{\text{fiber}}$ is the overlap integral in
Eq.~\ref{eq_overlap}, taken between the radiated field at position
$z$ and the fundamental (isolated) fiber mode. The factor of 2
accounts for collection from both fiber ends. The symmetry of the
geometry allowed us to choose symmetric
($\hat{\mathbf{x}}\times\mathbf{H}=0$) boundary conditions on the
$yz$-plane, as only $y$-polarized dipoles were considered.
Perfectly-matched layers (PMLs) were used around the domain limits
to simulate an open domain. Simulations ran until no field amplitude
could be detected in the domain. As in \cite{ref:Davanco}, $\etaPL$
oscillates with $z$, due to the back-and-forth power exchange
between the guide and the slab along the waveguide. The values of
$\etaPL$ reported below correspond to maxima obtained within the
computational window.

\subsection{Supermode analysis}
\label{subsection:supermode_analysis} We analyze the results from
the FDTD simulations in terms of the supermodes supported by the
coupler structure formed by the fiber and slab. This provides us
with insight into the collection mechanisms and ways to improve it,
as well as the ability to determine the possibility of observing
extinction of resonant input signals by a single dipole.  The
formalism we employ closely follows that used in our previous
work~\cite{ref:Davanco}.

Supermode field profiles and the respective complex propagation
constants $\beta_m$ are obtained with a finite-element based
eigenvalue solver, with a vectorial formulation. Supermode $m$'s
individual contribution to the total PL collection efficiency
$\etaPL$, considering one fiber channel, is
$\etaPLm{m}=f_m\cdot\Gamma_m / \Gamma = f_m\cdot\gamma_m$, where
$\Gamma_m$ is the supermode emission rate, and $\Gamma$ the total
emission rate. The fraction $\gamma_m$ is supermode $m$'s
spontaneous emission coupling factor ($\beta$-factor), which, since
emission in both $\pm z$ directions is equally likely, is such that
$0\leq\gamma_m\leq0.5$. The fiber mode fraction, $f_m$, is the
percentage of supermode $m$'s power that is transferred to the
output fiber mode. Assuming that the fiber is abruptly removed from
the slab at the end of the probing region, and that reflections at
the interface are small, $f_m$ may be approximated with an overlap
integral between the fundamental fiber mode and supermode $m$
~\cite{ref:Huang3,ref:Huang_integral_note}:
\begin{equation}
\f_m=\frac{ \text{Re}\left\{
\iint_{S}(\mathbf{e}_f\times\mathbf{h}_m^*)\cdot\hat{\mathbf{z}}\,dS
\,
\iint_{S}(\mathbf{e}_m\times\mathbf{h}_f^*)\cdot\hat{\mathbf{z}}\,dS\right\}
} { \text{Re}\left\{
\iint_{S}(\mathbf{e}_f\times\mathbf{h}_f^*)\cdot\hat{\mathbf{z}}\,dS
\right\} \, \text{Re}\left\{
\iint_{S}(\mathbf{e}_m\times\mathbf{h}_m^*)\cdot\hat{\mathbf{z}}\,dS \right\}
}. \label{eq_overlap}
\end{equation}
In this expression, valid for purely dielectric waveguides,
$\left\{{\mathbf{e}_m},{\mathbf{h}_m}\right\}$ and
$\left\{{\mathbf{e}_f},{\mathbf{h}_f}\right\}$ are the supermode and
fundamental fiber mode fields, respectively.  The supermode emission
rates $\Gamma_m$ is also obtained from the field profiles, according
to the expressions given in~\cite{ref:Davanco}.

We point out that most of the supermodes supported by the fiber and
membrane structure are leaky~\cite{ref:Snyder_Love}, i.e., power
confined in the area beneath the fiber leaks away from it as the
supermode propagates. The power leakage rate per propagation length
is related to the imaginary part of the calculated supermode's
complex effective index $n_{eff}$. The number of free-space
wavelengths necessary for the supermode amplitude to decay by a
factor~$0<\delta<1$ is $ N_{\delta}=-\Rneff\ln(\delta)/2\pi$, where
$\Rneff = \text{Im}\{n_\text{eff}\}^{-1}$, and an effective
supermode propagation length $L_{\text{eff},\delta}=N_\delta\lambda$
may be defined.

\subsection{Field Extinction}
For coherent, resonant excitation, the power detected at the output
fiber port is a result of the interference between the excitation
signal and the resonance fluorescence from the emitter (which are at
the same wavelength), and may be either higher or lower than the
detected power in the absence of the dipole. In order to determine
the variation in the transmitted power level due to the presence of
a single dipole, we make use of the quantum optics input-output
formalism of~\cite{vanEnk.pra.69.043813}, with which we obtain
operators for the multimode field for $z>z_0$, i.e., past the dipole
location:
\begin{eqnarray}
\mathbf{E}^{(+)}(z,t) = i\sqrt{2\pi}\sum_m
\sqrt{\frac{\hbar\omega}{4\pi
S_m}}\mathbf{e}_me^{-i(\omega t-\beta_mz)}\times \nonumber \\
\times\left[
\hat{a}^m_{in}(t-n_mz/c)+\sqrt{\Gamma_m}^*\sigma_-(t-n_mz/c)
\right]. \label{eq-e-field-op}
\end{eqnarray}
Here, $\sigma_-$ is the atomic lowering operator, $\hat{a}^m_{in}$
is (incident) supermode $m$'s input field annihilation operator,
$\mathbf{e}_m$ is the electric field distribution, $\beta_{m}$ the
propagation constant, $n_m$ the phase index, and $S_m=\text{Re} \{
\int_S\,dS( \mathbf{e}_m\times\mathbf{h}^*_m )\cdot\hat{\mathbf{z}}
\}$, with $S$ the $xy$ plane. The expression in brackets is a
well-known result of the input-output formalism, with explicit input
(or "free") fields and radiated ("source")
contributions~\cite{vanEnk.pra.69.043813}, expanded in terms of
supermodes. The field operators are then inserted in the fiber mode
power operator~\cite{Davanco2},
\begin{eqnarray}
\hat{F}=\left\{
\int_SdS(\mathbf{E}^{(-)}\times\mathbf{h}_f)\cdot\hat{\mathbf{z}}
\int_SdS(\mathbf{e}^*_f\times\mathbf{H}^{(+)})\cdot\hat{\mathbf{z}}+ \right. \nonumber \\
\left.\int_S\,dS(\mathbf{H}^{(-)}\times\mathbf{e}_f)\cdot\hat{\mathbf{z}}
\int_SdS(\mathbf{h}^*_f\times\mathbf{E}^{(+)})\cdot\hat{\mathbf{z}}
\right\}S^{-1}_f, \label{eq-F-op}
\end{eqnarray}
where $\mathbf{e}_f$ and $\mathbf{h}_f$ are the fiber mode electric
and magnetic field distributions, $S_f=\text{Re} \{ \int_S\,dS(
\mathbf{e}_f\times\mathbf{h}^*_f )\cdot\hat{\mathbf{z}} \}$. In
short, the fiber mode power operator allows us to determine the
total photon flux coupled into the output fiber mode, based on
multimode field operators that describe the coherent interference
between the incident ('free') and emitted, resonance fluorescence
('source') supermode waves; the operator is the quantum optics
equivalent to the overlap integral in Eq.~(\ref{eq_overlap}),
between the total field at a position z along the waveguide and the
optical fiber mode. This corresponds to the power coupled into the
output fiber at the end of the coupling region, assuming an abrupt
transition and small reflections at the interface~\cite{ref:Huang3}.
Considering coherent, steady-state, multimode field excitation, the
photon flux $F$ at the output fiber (normalized to the input field
power) is found to be
\begin{eqnarray}
F&=&
\hbar\omega\Real{\sum_{m,m'}\sqrt{f_m}\sqrt{f_{m'}}e^{i(\beta_{m'}-\beta_m)(z-z_0)}\right.\times\label{eq:Flux}\\
&\times&\left.\left[\Bmm+\frac{\sqrt{\Gamma_m}
\sqrt{\Gamma_{m'}}^*\zeta- \frac{\Gamma}{2}
\left(B_m^*\sqrt{\Gamma_{m'}}^*\xi+B_{m'}\sqrt{\Gamma_{m}}\xi^*\right)}
{\left(\frac{\Gamma}{2}\right)^2+2\zeta}\right]},\nonumber
\end{eqnarray}
with $\xi=\displaystyle\sum_{m}\sqrt{\Gamma_m}B_m$,
$\zeta=\displaystyle\sum_{m,m'}\Real{
\sqrt{\Gamma_m}\sqrt{\Gamma_{m'}}^*B_m^*B_{m'} }$. In this
expression, $B_m$ is the complex amplitude of the $m$-th supermode
incident on the dipole. The magnitude of $B_m$ is determined by the
manner with which the fiber is brought into contact with the slab.
For instance, for abrupt contact (e.g., very short transition
regions in Fig.\ref{fig-probing-scheme}), it approaches
$|f_m|^{1/2}$, where $f_m$ is the fiber-mode fraction. Longer
transition regions could lead to a power distribution among the
excited supermodes different from that obtained with the fiber-mode
fractions. The phase of the $B_m$ coefficients at the dipole
position is determined through the supermode propagation constants
$\beta_m$, assuming all modes are in-phase at the start of the
coupler region.

To gain some insight into the mechanisms at play at resonant
excitation, we consider a situation in which only one supermode of
the fiber and slab structure may be accessed: in Eq.(\ref{eq:Flux}),
we set all fiber-mode fractions $f_m$ and incident supermode
amplitudes $B_m$ to be null except for those of an arbitrary $M$-th
supermode (i.e., $f_{m\neq M}=0$, $B_{m\neq M}=0$). We furthermore
make the assumption that the excitation signal drives the transition
far from saturation, so that $\zeta/\Gamma^2\ll1$ and the
denominator of the second term in brackets becomes unity. In this
case, the power detected at the output fiber is proportional to
$1-4\gamma_M(1-\gamma_M)$. Thus the magnitude of optical field
extinction by a single dipole is determined by the supermode
spontaneous emission coupling factor, and is complete when
$\gamma_M=0.5$.

\section{Fluorescence Collection Efficiency}
\label{section 2} Two configurations were analyzed, for operation at
visible ($\lambda=600\nm$) and near-infrared ($\lambda=1300\nm$)
wavelengths. For the visible range, a 130~nm thick SiN membrane
(refractive index $n_\text{SiN}$ = 2.0) and 400~nm diameter single
mode fiber taper waveguide were considered. In the near-infrared
case, our model consists of a 160~nm thick Si membrane (refractive
index $n_\text{Si}$ = 3.505) and a $1\mum$ diameter fiber taper,
which supports a well-confined and a near-cutoff mode. The first
configuration models a system suitable for probing visible
wavelength emitters such as single molecules or CdTe/ZnSe
nanocrystal quantum dots attached to the SiN membrane, while the
second would be suitable for infrared emitters, such as PbS and PbSe
nanocrystal quantum dots~\cite{ref:Wise_ACR00}. In both cases, the
emitters are considered to be embedded in a 20 nm thick, purely
dielectric host film on top of the SiN or Si membranes, as shown in
Fig.\ref{fig-probing-scheme}(b). The host film refractive index
$n_{host}$ is allowed to vary between 1.0 and 1.7, a range that
includes typical values for possible organic (e.g., PMMA) or
inorganic (e.g., silica) transparent host materials. The emitters
are modeled as electric dipoles oriented in the $y$-direction. While
this consideration limits our analysis to a best-case scenario for
emitters with random dipole orientation, it may be well suited to
model certain organic crystal guest-host systems, where emitting
molecules embedded in the host crystal tend to align in specific
orientations. For instance, in the guest-host system presented
in~\cite{ref:Pfab}, consisting of Terrylene molecules in a
crystalline $p$-terphenyl host -a system fit for molecular quantum
optics~\cite{ref:Moerner}-, the guest molecules have been shown to
display dipole moments perpendicularly oriented to the substrate.

\begin{figure}[h]
\centerline{\includegraphics[width=13.0cm]{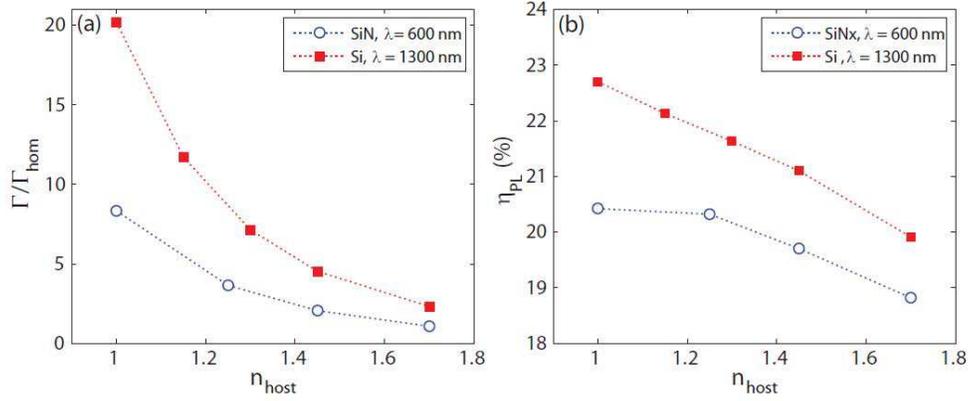}}
\caption{(a)Maximum total spontaneous emission rate enhacement
$\Gamma/\Gamma_\text{hom}$, where $\Gamma_\text{hom}$ is the
spontaneous emission rate of a dipole in a homogeneous dielectric
medium of refractive index $n_\text{host}$. (b) PL collection
efficiencies $\etaPL$, including both fiber ends, for $y$-polarized
dipoles in the SiN ($\lambda=600\nm$) and Si ($\lambda=1300\nm$)
membrane configurations, as functions of the host layer refractive
index, $n_\text{host}$. Results calculated with finite difference
time domain simulations.} \label{fig-etaPL}
\end{figure}

Figure \ref{fig-etaPL}(a) shows the FDTD-calculated total
spontaneous emission rate $\Gamma$, for a $y$-polarized dipole in
the SiN ($\lambda=600\nm$) and Si ($\lambda=1300\nm$) membrane
configurations, as functions of the host layer refractive index,
$n_\text{host}$. All rates are normalized to the rate for a dipole
in a homogeneous space of index $n_\text{host}$
($\Gamma_\text{Hom.}$) and thus correspond to the spontaneous
emission enhancement (Purcell) factor. Purcell factors as high as 20
and 9 respectively are observed for $n_\text{host}=1.0$ in the Si
and SiN cases, and decrease with increasing host index. Simulations
with progressively denser meshes were used to verify that these
results converged to within at least 3~\%.

In Fig.~\ref{fig-etaPL}(b), the corresponding maximum PL collection
efficiencies $\etaPL$ including collection from both fiber ends are
plotted. As mentioned earlier, since $\etaPL$ oscillates in $z$ due
to power exchange between the fiber and slab, the plotted values
correspond to the maximum efficiency within the computational window
(with dimensions $>6$ wavelengths). It is apparent that collection
efficiencies are above 18~$\%$ for all $n_\text{host}$ in both Si
and SiN cases, with maxima of $22.8~\%$ and $20.5~\%$ at
$n_\text{host}=1.0$ respectively, and decreasing for higher values.

\section{Supermode Analysis}
\label{section 3} 
\begin{figure}[t]
\centerline{\includegraphics[width=13.0cm]{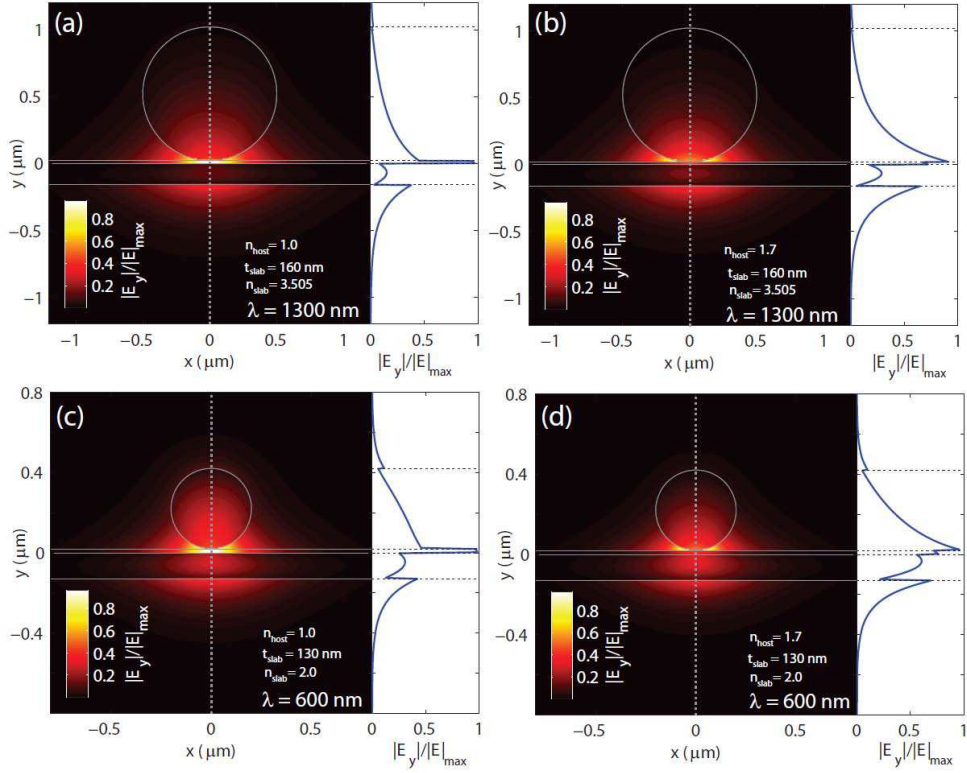}}
\caption{Amplitude of the major electric field component ($E_y$) of
laterally bound gap modes (normalized to the maximum electric field
amplitude, $|\mathbf{E}|_\text{max}$) for the (a)-(b) Si slab
configuration ($\lambda=1300\nm$) with (a) $n_\text{host}=1.0$ and
(b) $n_\text{host}=1.7$; (c)-(d) SiN configuration
($\lambda=600\nm$) with (c) $n_\text{host}=1.0$ and (d)
$n_\text{host}=1.7$. In all cases, $t_{\text{host}}$ = 20 nm. Line
plots show $|{E_y}|/|\mathbf{E}|_\text{max}$ on the $x=0$ plane
(dotted line in the contour plots).} \label{fig-fields}
\end{figure}

A modal analysis of the the coupler structures
formed by the fiber and membranes (cross section shown in
Fig.\ref{fig-probing-scheme}(b)) reveals the existence of laterally
confined supermodes, such as depicted in Figs. \ref{fig-fields}(a)
and (c), with a predominantly $y$-polarized electric field strongly
concentrated in the host material layer. The strong field
concentration translates into an enhanced dipole coupling strength,
and leads to the spontaneous emission enhancement reported in
Section~\ref{section 2}. Group velocities are not significantly low,
so the density of modes does not play an important role in emission
enhancement, contrasting with the situation in slow-light photonic
crystal waveguides~\cite{ref:rao,ref:Lecamp}.

\begin{figure}[h]
\centerline{\includegraphics[width=13.0cm]{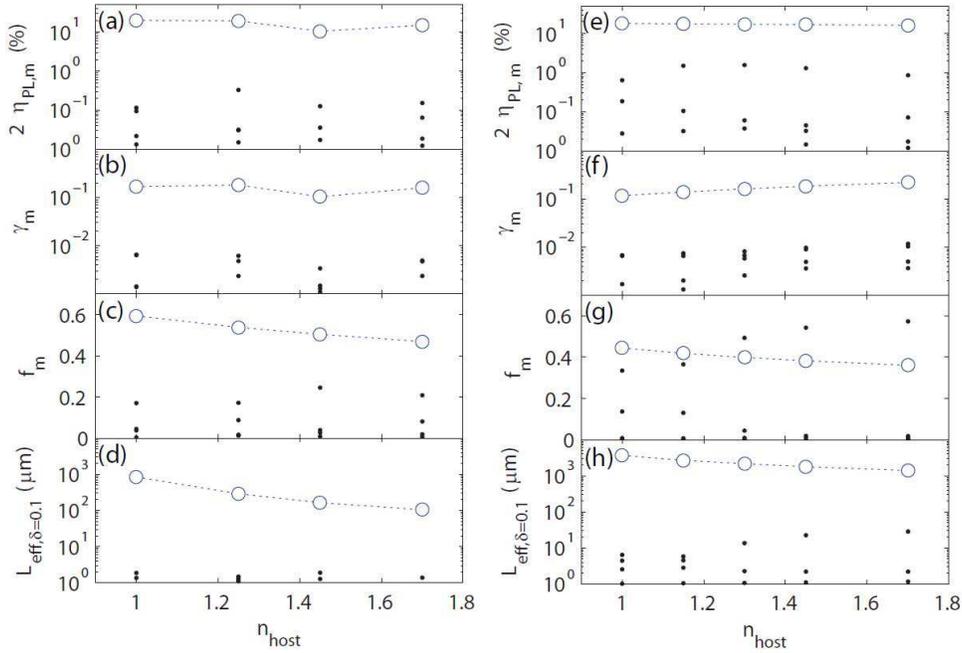}}
\caption{Supermode contributions to the total PL collection
efficiency ($\eta_\text{PL,m}$), modal spontaneous emission coupling
factors ($\gamma_m$), fiber mode fractions $\f_m$ and effective
supermode lengths $L_{\text{eff},\delta=0.1}$ for the (a)-(d) Si
slab, $\lambda=1300\nm$ and (e)-(h) SiN, $\lambda=600\nm$ systems.
Circles: main supermode; dots: secondary supermodes }
\label{figeta_m}
\end{figure}

In both, Si and SiN, systems, the strong field concentration
originates in the large index steps between the membrane and host
layer. Such gap modes are the main contributors to the total
collection efficiency, $\etaPL$. This is apparent in
Figs.~\ref{figeta_m} (a) and (e), where the individual
contributions, $\eta_\text{PL,m}$, to the total collection
efficiency, $\etaPL$, are plotted. The contributions of all
secondary supermodes (black dots in Fig.\ref{figeta_m} (a) and (e))
are at least an order of magnitude smaller than the main ones.
Despite providing small individual contributions, secondary
supermodes altogether make up for a large portion of the total
collected power. It is important to note that, even though many of
the secondary supermodes are indeed gap modes, poor lateral
confinement causes these to exhibit large effective areas, and
therefore low emission rates $\Gamma_m$. Correspondingly, low
emission coupling factors $\gamma_m$ ($\gamma_m=\Gamma_m/\Gamma$,
with $\Gamma$ the total spontaneous emission rate) are observed in
Figs.~\ref{figeta_m}(b) and (f).  For the main modes, $\gamma_m>
10~\%$, so these gap modes carry a considerable percentage of the
total spontaneous emission.

The oscillation of the total efficiency $\etaPL$ as a function of
$z$, mentioned in Section~\ref{section 2}, can be traced to the
beating of the two main contributing supermodes. The beat length is
given by $L_{\text{z}}=2\pi/(\Delta \beta)$, where $\Delta \beta$ is
the difference between the propagation constants. For the Si,
$\lambda=1.3\mum$ configuration, the beat length varies between
$4.78\mum$, for $n_\text{host}=1.0$ and $3.89\mum$, for
$n_\text{host}=1.7$. In the SiN, $\lambda=0.6\mum$ case, it varies
between $2.72\mum$, for $n_\text{host}=1.0$ and $2.17\mum$, for
$n_\text{host}=1.7$. In order to maximize the collected power,
therefore, control of the interaction length on the scale a few
microns is desirable.

At the same time, short interaction lengths are desirable because
all supermodes exhibit some degree of lateral power leakage due to
imperfect field confinement. Figures~\ref{figeta_m}(d) and (h) show
the effective lengths $L_{\text{eff},\delta}$, defined in
Section~\ref{subsection:supermode_analysis}, necessary for 10~\%
supermode amplitude decay, for the Si and SiN case respectively. It
is apparent that in both cases the main supermodes have considerably
longer (more than two decades) effective lengths, as expected, in
view of their lower lateral power leakage rate. For the Si structure
at $\lambda=1300$ nm, the necessary length for the main supermode to
decay by 10~$\%$ varies between 3650$\mum$ and 1400$\mum$ for the
host index range considered. In the SiN case at $\lambda=600\nm$,
the 10~$\%$ decay length ranges between 1840$\mum$ and 225$\mum$.
For interaction lengths of only a few microns, the main supermode
contributions in Figs.~\ref{figeta_m} (a) and (e) may thus be taken
as lower bounds for the total achievable collection efficiency.

\section{Resonant Extinction Spectroscopy }
\label{section 4} 
\begin{figure}[htfb]
\centerline{\includegraphics[width=13.0cm]{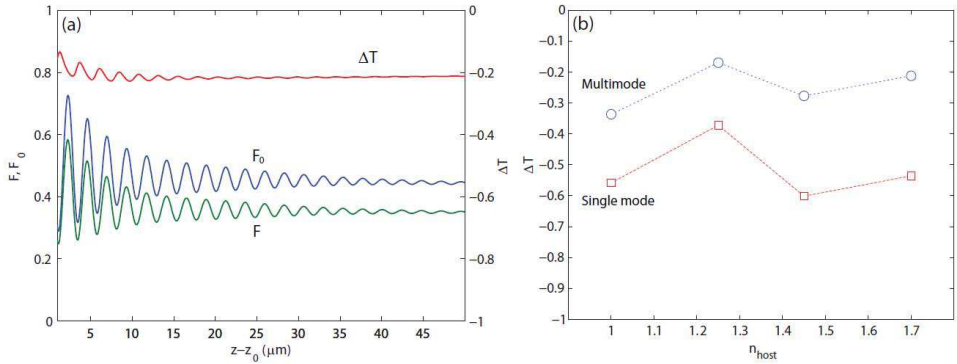}}
\caption{(a) Normalized, off- and on-resonance transmission ($F_0$
and $F$) and contrast $\Delta T=(F-F_0)/F_0$ as functions of
separation from a single, $y$-oriented dipole at $z_{0}$. The dipole
is embedded in a host material with $n_\text{host}=1.7$ on top of a
130~nm thick SiN membrane, and emits at $\lambda=600 \nm$. (b)
Achievable transmission contrast $\Delta T$ as a function of the
host film index $n_\text{host}$. Squares: results obtained assuming
dipole excitation with the main supermode only; circles: assuming
multimode excitation (see text for details).} \label{fig:extinction}
\end{figure}

The SiN membrane configuration offers positive
prospects for performing extinction-based single emitter
spectroscopy measurements with a coherent resonant excitation
signal~\cite{gerhardt.prl.033601}. The main contributing supermode
has a modal spontaneous emission coupling factor $\gamma_m>0.1$ with
maximum of 0.18 for $n_\text{host}$=1.25 (see
Fig.~\ref{figeta_m}(f)); if excited alone, the field extinction at
the output fiber could be higher than 37~$\%$, with a maximum of
60~$\%$ for $n_\text{host}=1.25$. For an abrupt transition, however,
as explained above, additional supermodes would be excited, with
efficiencies given by the fiber-mode fractions $f_m$. Such
additional supermodes are not strongly extinguished by the dipole,
having $\gamma_m<0.006$, and thus limit the achievable overall
extinction at the output fiber. To estimate output fiber power
extinction, we use Eq.~(\ref{eq:Flux}) with the three highest
contributing supermodes, considering $n_\text{host}=1.7$. The
percentage of fiber mode power carried by these three supermodes
altogether amounts to 75.5~$\%$ (see Fig.~\ref{figeta_m}(g)). We
assume that the remainder of the power is completely transferred to
the output fiber, a worst-case scenario which implicitly assumes
that none of this power leaks away from the fiber, or is reflected
by the dipole. Figure~\ref{fig:extinction} (a) shows the calculated
optical flux at the output fiber for light on- and off-resonance
($F$ and $F_0$) with a $y$-oriented dipole in a host material of
index 1.7, as well as the transmission contrast through the fiber,
${\Delta}T = (F-F_0)/F_0$. In the figure, the horizontal axis is the
distance from the dipole to the end of the coupler region, as
indicated in Fig.~\ref{fig-probing-scheme}, and $z_0=1\mum$.  The
oscillations in $F$, $F_0$ and $\Delta T$ along $z$ evidence the
back-and-forth power transfer between fiber and slab, and the
amplitude decay is indicative of both the interference and leaky
nature of the supermodes involved. It is apparent that, for $z-z_0
\gg 0$, the extinction level may reach $\approx 20~\%$.

Figure~\ref{fig:extinction}(b) shows the achievable extinction ratio
at $z-z_0=50\mum$, under the same assumptions, for all host indices
considered (circles). We also plot the extinction levels achievable
under the single supermode assumption (squares), which makes
explicit the degradation of extinction due to the excitation of
additional supermode waves. The highest achievable extinction may be
as much as 34~\% for $n_\text{host}=1.0$, for which the Purcell
enhancement is maximum and considerably stronger than for
$n_\text{host}=1.7$. This result suggests that, as long as the input
signal power is sufficiently below the transition saturation power,
the Purcell effect does not influence the achievable extinction
level. This is to be expected, given that the aforementioned
expression for extinction in the case of single mode excitation and
collection, $F\sim 1-4\gamma_m(1-\gamma_m)$, only depends on the
modal spontaneous coupling factor $\gamma_m$. Purcell enhancement
will influence the power at which dipole saturation occurs, and may
thus have implications in obtaining better detection signal-to-noise
ratios.

Similar performance may in principle be achieved with the Si system,
for an appropriate set of parameters. The situation, however, is
less favorable for the parameters considered here: in
Fig.~\ref{figeta_m}(c), for $n_\text{host}>1.2$, the main supermode
fiber fraction is surpassed by that of the second mode, which is
only weakly affected by the dipole ($\gamma_m<0.01$ in
Fig.~\ref{figeta_m}(b)). Assuming that the second supermode is
completely transmitted, that 40~$\%$ of the power in the fiber is
carried by the main supermode (see Fig.~\ref{figeta_m}(c)), and that
the latter is extinguished by 31~$\%$ ($\gamma_m=0.22$,
Fig.~\ref{figeta_m}(b)), the maximum achievable overall extinction
would be only 12~$\%$. Although lower than in the SiN cases studied
above, such an extinction level is still quite reasonable for
spectroscopy purposes, and is compatible with experimentally
observed levels using an NSOM tip~\cite{gerhardt.prl.033601} and a
solid immersion lens~\cite{Wrigge.Nature.Phys.4.60}.

\section{Discussion}
\label{section 5} In slot waveguides consisting of two high
refractive index channels separated by a small
gap~\cite{Almeida.OL.04}, strong field concentration in the gap
results from the continuity of the normal electric displacement
vector component at dielectric interfaces, with the field
concentration increasing for increasing refractive index
discontinuity. This behavior is evident in
Figs.~\ref{fig-fields}(a),(b) and (c),(d), which depict the
intensity of the electric field in the host region, for host film
refractive indices $n_\text{host}=1.0$ and $n_\text{host}=1.7$, for
both the Si and SiN configurations. The decrease in field
concentration for larger indices is apparent in the Purcell factor
decrease for increasing host material index $n_\text{host}$
(Fig.~\ref{fig-etaPL}(b)). However, even in situations where the
Purcell factor is small, gap supermodes have relatively high modal
spontaneous emission coupling factors ($\gamma_{m}$) and efficient
coupling to the fiber mode ($f_{m}$) (see Fig.~\ref{figeta_m}),
which creates the conditions for performing extinction-type,
resonant spectroscopy on individual emitters.

We point out that the cases in which the highest Purcell
enhancements are observed, in which $n_\text{host}=1.0$, would be
challenging or not achievable in practice. These situations, which
otherwise provide upper bounds for the achievable enhancement, imply
the absence of a host material supporting the emitter, or, in the
best case, the ability to produce a 20 nm thick host layer of
extremely low refractive index material, for instance an aerogel. It
is also worthwhile noting that, although semiconductor nanocrystal
quantum dots are composed of high refractive index materials,
sufficiently small nanocrystals may not constitute a significant
disturbance to the environment, that could lead to large deviations
from our calculated results. Evidence of this can be found
in~\cite{ref:Rakher_arxiv}, where small ($< 5$ nm) PbS nanocrystal
quantum dots were shown to not considerably affect the modes of a
high quality factor microresonator. A similar conclusion may be
drawn from the simulations involving diamond nanocrystals in
nanocavities reported in~\cite{ref:McCutcheon}.

We reiterate that the results presented here are best-case
estimates, obtained for vertical dipole moments aligned with the gap
mode field. Horizontal dipole moments are expected to radiate at
lower rates. In the case of $z$-dipoles, gap waves are generated at
lower rates because the dipole moment is aligned with the minor
supermode field component, $E_z$. If located at the $x=0$ plane, an
$x$-dipole is completely uncoupled from gap modes, producing,
rather, anti-symmetric ($\hat{\mathbf{y}}\times
\mathbf{E}|_{x=0}$=0) supermode waves with a major $x$-field
component. These do not offer the same strong field concentration as
gap modes, due to the continuity of the electric field across the
gap.

As in the situation studied in~\cite{ref:Davanco}, relating to PL
collection from slab-embedded dipoles, lateral modal confinement
plays an important role. Despite the high $\gamma_{m}$ factors of
the main gap supermodes, a significant portion of the dipole power
is emitted into supermodes with very poor lateral confinement, which
in a short distance along $z$ propagate away from the fiber and are
not collected. Thus, the length of the interaction region must be
minimized for a maximized collection. In addition, enforcing strong
lateral confinement via the formation of a channel, rather than a
membrane, may significantly enhance collection efficiency. For
instance, by replacing the SiN membrane studied above with a channel
waveguide of the same thickness, but with a width of 600~nm and with
a host material of index $n_\text{host}=1.45$, a collection
efficiency of $\approx49~\%$ may be achieved with a Purcell factor
of $\approx2$. Additionally, as in ~\cite{Davanco2}, a channel
waveguide structure may also benefit resonant excitation
experiments, not only with an enhanced resonance fluorescence
collection efficiency, but also with higher extinction levels, or
even enhancement of the output power relative to the transmission
level without the dipole. In this case, tapered fiber and channel
waveguides may be designed to form a phase-matched directional
coupler, with high $\gamma_{m}$ supermodes. The waveguide could
furthermore be designed to allow efficient power transfer from the
fiber to a single supermode, so that the single-mode situation
mentioned in Section~\ref{section 4} may be achieved. In this case,
provided the single supermode's $\gamma_{m}$ is large, high field
extinction may be achieved independent of the coupler length or
emitter position along its length~\cite{Davanco2}. For the channel
waveguide situation just described, $\gamma_m=0.215$ for the main
supermode, which would lead to a single-mode extinction of 67~\%.

Finally, we point out that the recently reported plasmonic
laser~\cite{ref:Oulton}, composed of a cylindrical CdS nanowire
placed on top of a Ag substrate, with a thin MgS separator (the
cross section closely resembles that in
Fig.~\ref{fig-probing-scheme}(b)) supports gap modes similar to
those studied here. While here the supermodes are hybrids of fiber
and slab modes, those in~\cite{ref:Oulton} are hybrids of the
cylinder and surface plasmon modes. Although hybrid plasmonic
waveguides offer substantially stronger Purcell enhancement for
emitters located in the MgS spacer or in the nanowire, the
supermodes suffer from high propagation losses associated with the
surface plasmon. Indeed, supermode propagation lengths (i.e., the
required length for the mode power to drop in half) quoted in
~\cite{ref:Oulton} are on the order of 10 wavelengths. In the
present case, supermode effective lengths are two to three orders of
magnitude longer, since the waveguides are purely dielectric, and
lateral power leakage is small.

\section{Conclusions}
\label{section 6}

We have performed detailed electromagnetic simulations that indicate
that a tapered optical fiber waveguide may be an efficient
photoluminescence collection probe for individual dipoles placed on
the surface of dielectric membranes. High maximum collection
efficiencies ($>20~\%$) are predicted, and are related to hybrid
supermodes of the composite fiber-slab structure, which present
strong field concentration in the gap between slab and fiber. These
supermodes are akin to the gap modes supported by slotted
waveguides~\cite{Almeida.OL.04}. Our results also indicate that this
probing configuration may be used to perform resonant spectroscopy
of single emitters on membrane surfaces, and we predict that an
extinction of at least 20~\% of a fiber-coupled resonant excitation
signal is achievable.

\section*{Acknowledgement}
This work has been supported in part by the NIST-CNST/UMD-NanoCenter
Cooperative Agreement.


\begin{thebibliography}{10}
\bibitem{ref:Huang_integral_note}
Following Ref.[19], the fiber mode fraction, Eq.(\ref{eq_overlap}),
would be
  given by the expression $f_m = \langle f |m \rangle \langle m |f
  \rangle\left(\langle f |f \rangle \langle m |m \rangle\right)^{-1}$, where
  $\langle f |m \rangle=\iint\left( \mathbf{e}_f \times \mathbf{h}_m^* +
  \mathbf{h}_f \times \mathbf{e}_m^* \right)\cdot\hat{z}\,dS/4$. Considering no
  reflections at the interface between the isolated fiber and the contact
  region, (i.e., the field just after the interface is identical to the
  incident, foward propagating, field), Eq.(\ref{eq_overlap}) gives the same
  result.
\end{thebibliography}
\end{document}